\documentstyle[aps,twocolumn,epsf]{revtex}
\begin{document}
\title{Unambiguous State Discrimination of Coherent States with Linear Optics: Application to Quantum Cryptography}
\author{S.J. van Enk\\
Bell Labs, Lucent Technologies,
Room 2C-401\\
600-700 Mountain Ave,
Murray Hill NJ 07974}
\maketitle

\begin{abstract}We discuss several methods for unambiguous state discrimination of $N$ symmetric coherent states using linear optics and photodetectors. One type of measurements is shown to be optimal in the limit of small photon numbers for any $N$.  For the special case of $N=4$ this measurement can be fruitfully used by the receiving end (Bob) in an implementation of the BB84 quantum key distribution protocol using faint laser pulses. In particular, if Bob detects only a single photon the procedure is equivalent to the standard measurement that he would have to perform in a single-photon implementation of BB84,
if he detects two photons Bob will unambiguously know the bit sent to him in 50\% of the cases without having to exchange basis information, and if three photons are detected, Bob will know unambiguously which quantum state was sent.\end{abstract}
\medskip
\section{Introduction}
The simple fact that nonorthogonal quantum states cannot be perfectly distinguished lies at the basis of Quantum Cryptography and Quantum Key Distribution (QKD): An eavesdropper is not able to determine with 100\% certainty which quantum state was sent if the alternatives were chosen from a nonorthogonal set of states. Among the various strategies the eavesdropper might use to gather at least some information are measurements that maximize her probability of guessing the state correctly. Alternatively, she might decide to perform a measurement that allows her with some nonzero probability to learn unambiguously which state was sent, at the cost of sometimes getting an inconclusive result. The concept underlying this type of measurements, unambiguous state discrimination (USD), was first introduced for the case of two nonorthogonal states by Ivanovic in 1987 \cite{ivanov} and a bound on the maxium achievable success probability was found shortly afterwards \cite{dieks},
\begin{equation}
P_D^{(2)}=1-|\langle \psi_0|\psi_1\rangle|,
\end{equation}
if $|\psi_{0}\rangle$ and $|\psi_{1}\rangle$ are the two states to be distinguished.

In general, a USD measurement with a nonzero probability to succeed exists for any number of states provided they are linearly independent \cite{cheflesa}. For the special case of sets of $N$ {\em symmetric} states, i.e., states $\{|\psi_i\rangle\}_{i=1\ldots N}$ that can be written as $\{ U^{i-1} |\psi_1\rangle \}_{i=1\ldots N}$ for some fixed unitary operation $U$ such that $U^{N}|\psi_1\rangle=|\psi_1\rangle$,  the optimum USD probabilities were found in \cite{chefles}.
Eavesdropping attacks using such USD measurements were analysed in \cite{brandt,norbert} and USD experiments on polarization were reported on in \cite{expusd}.
Here we will consider linear optics implementations of USD measurements on symmetric coherent states.
In particular, we show that even without entangled measurements and Quantum Non-Demolition (QND) measurements
a USD measurement with near-optimum success probablity is possible using just beamsplitters, photodetectors and feedback.

The motivation for this is as follows:
In virtually all experimental implementations of the BB84 protocol \cite{bb84} the signal states are weak laser pulses (see for instance \cite{exp}) rather than the single photons envisaged in the original protocol. One disadvantage is that an eavesdropper may exploit the presence of multiple photons in some of the signals to gain more information and/or remain undetected, as has been discussed in several papers \cite{multi}. We show here how the legitimite users too can in fact make use of multiple photons by improving upon the standard measurement that is used in current implementations, which indeed is tailored for the single-photon implementation.
If the quantum information is encoded in the phase of the signal a strong reference pulse is needed to provide a phase reference. The quantum states sent may then be assumed to be pure coherent states (relative to the reference pulse, see \cite{enkfuchs}) with known amplitude $\alpha$ (assumed to be real) but unknown phases that may take one of four possible values, $0,\pi/2,\pi,3\pi/2$. As pointed out in \cite{norbert}, if polarization is used instead of phase, the signal states are mixed states, not coherent states. Yet, as detailed below, an important part of the following applies to both polarization and phase encoding.

In Section \ref{rules} we formulate the problem solved in this paper. In Section \ref{results} we discuss the solutions for USD of $N$ symmetric coherent states for $N=2,3,4$ and $N>4$ separately. The most detailed exposition is given for the case $N=4$, the case that applies to QKD. 
\section{Formulation of the problem}\label{rules}
Suppose one has one copy of a coherent state $|\alpha e^{i\phi}\rangle$ with known amplitude $\alpha$ but an unknown phase $\phi$ that may have one of $N$ values, $\phi_k=2\pi k/N$ with $k=0,1\ldots (N-1)$. The task is to unambiguously discriminate between these $N$ values. Since the set of states $\{ |\alpha e^{i\phi_k}\rangle, k=0,1,\ldots (N-1)\}$ is linearly independent there are in principle measurements possible that allow one to accomplish unambiguous state discrimination (USD) with some nonzero probability. An optimum protocol and the corresponding maximum probabilities $P_D^{(N)}$ have been derived in \cite{chefles}:
Defining probability amplitudes $|c_k|^2$ for $k=0\ldots N-1$ by
\begin{equation}
|c_k|^2=\frac{1}{N}\sum_{j=0}^{N-1} e^{-2\pi ijk/N} e^{|\alpha|^2(e^{2\pi i j/N}-1)},
\end{equation}
$P_D^{(N)}$ is determined by the smallest of these amplitudes, according to
\begin{equation}\label{min}
P_D^{(N)}=N\min_{k=0\ldots N-1} |c_k|^2.
\end{equation}
The question considered here is what USD probability $P_{BS}^{(N)}$ can one reach while using just beamsplitters, photodetectors, feedback and the ability to produce known coherent states of any amplitude?

To be more precise, we allow the following operations and measurements:
\begin{enumerate}
\item Unitary operations: take two coherent states $|\beta\rangle$ and $|\gamma\rangle$ (that may be known or derived from the initial unkown state $|\alpha e^{i\phi}\rangle$) and split them on a beamsplitter  to get two new coherent states as output modes, with amplitudes 
$t\beta + r\gamma$ and $r\beta + t\gamma$, in terms of the transmission and reflection coefficients $t,r$ of the beamsplitter.
\item Measurements: take a coherent output state and measure
whether it contains photons or not, with a photodetector of efficiency $\eta$.
\item
Feedback: some of the light may be split off to a delay line so that subsequent measurements and unitary operations may depend on previous measurement outcomes.

\end{enumerate}
Neither Quantum-Non-Demolition (QND) measurements nor entangled measurements are considered here, as those
are far more difficult to perform with present-day technology.

\section{Results}\label{results}
\subsection{Eliminating one phase value}\label{elim}
A primitive to be used later on is a simple beamsplitter setup that allows one with a finite probability to eliminate one particular value of the phase of a coherent state with known amplitude.
Take the state $|\alpha e^{i\phi}\rangle$ with unknown phase
and combine it on a beamsplitter with the known coherent state $|\beta e^{i\phi_0}\rangle$ with $\beta=-t/r\alpha$ and $\phi_0$ the phase we wish to eliminate. By taking the limit of $t\rightarrow 1$ and $r\rightarrow 0$ one will get (besides a useless output that is discarded) a coherent state  $|\alpha (e^{i\phi}-e^{i\phi_0})\rangle$. (Note this corresponds to a displacement operation.) If a photon is detected in this state $\phi$ cannot have the value $\phi_0$.  The probability of such a detection event is 
\begin{equation}\label{P2}
P_{\eta}(\alpha,|\phi-\phi_0|)\equiv 1-e^{-\eta|\alpha|^2|e^{i\phi}-e^{i\phi_0}|^2},
\end{equation}
which depends on the actual (unknown) phase $\phi$. We included here the quantum efficiency $\eta$, defined as the probability of the photodetector to detect the presence of a photon. The effect of a finite efficiency is simply to effectively reduce the amplitude of the coherent state $\alpha$ by a factor $\sqrt{\eta}$. In the following we present results for perfect photodetectors only and the results for imperfect photodetectors can be obtained by subsituting $\alpha\rightarrow \sqrt{\eta}\alpha$.  
\subsection{$N=2$}\label{N=2} 
This case is straightforward and known. It basically is a measurement in a particular ``basis'' specified by two phase values $\pi$ apart.
Take the state $|\alpha e^{i\phi}\rangle$ and put it on a 50/50 beamsplitter (i.e., $t=1/\sqrt{2}$ and $r=i/\sqrt{2}$) with the state $|i\alpha\rangle$, such that the two output modes are coherent states with amplitudes $|\alpha (e^{i\phi}\pm 1)/\sqrt{2}\rangle$, and measure photons on both outputs. Since $\phi$ is either 0 or $\pi$ the probability to detect a photon is 0 in one output mode, and $1-e^{-2|\alpha|^2}$ in the other. If a photon is detected one can eliminate one of the two possible values of $\phi$, and thus the probability of USD with beamsplitters is 
\begin{equation}\label{2}
P_{BS}^{(2)}=1-e^{-2|\alpha|^2},
\end{equation}
which is in fact equal to the optimum value $P_D^{(2)}$.

Alternatively one could use the phase elimination scheme from Section~\ref{elim} to find the same optimum value. In fact, applying that method allows one to perform an optimal USD on two coherent states with arbitrary phases $\phi_0$ and $\phi_1$. Namely, split the original unknown state in two equal parts $|\alpha e^{i\phi}/\sqrt{2}\rangle$ and perform the two elimination measurements to eliminate $\phi_0$ or $\phi_1$. The probability to eliminate one of the two is
\begin{equation}
P_{BS}^{(2a)}=1-e^{-|\alpha|^2|e^{i\phi_0}-e^{i\phi_1}|^2/2},
\end{equation}
which is optimal and which reduces to (\ref{2}) for $\phi_0=\phi_1+\pi$.
\subsection{$N=3$}\label{N=3}
\begin{figure}\leavevmode
\epsfxsize=8cm \epsfbox{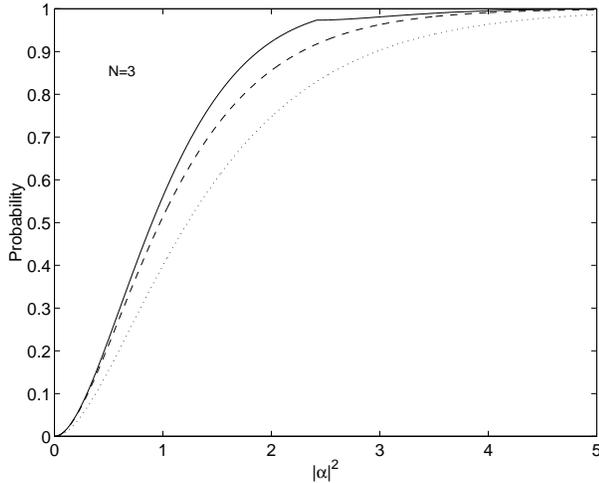} \caption{USD for $N=3$ symmetric coherent states. The solid curve gives the optimum success probability $P_D^{(3)}$, the dotted curve corresponds to a simple beamsplitter measurement $\tilde{P}_{BS}^{(3)}$, and the dashed curve describes a more complicated version of the beamsplitter setup $P_{BS}^{(3)}$.}
\label{3} \end{figure}

In order to analyse the case $N=3$ we first consider a simple USD set-up with a reasonably high probability of success. First split the original unknown coherent state into 3 identical copies 
$|\alpha e^{i\phi}/\sqrt{3}\rangle.$ Then use each of these three states in a phase elimination measurement setup of the type discussed in Section~\ref{elim}.
With the help of (\ref{P2}) the probability to eliminate two out of three phases is 
\begin{equation}
\tilde{P}_{BS}^{(3)}=\left(1-e^{-|\alpha|^2}\right)^2,
\end{equation}
since $|e^{i\phi_k}-e^{i\phi_l}|^2=3$ for all 3 possible pairs of different $k,l$.
This function is plotted in Fig.~\ref{3}.

One can improve upon this simple strategy by splitting the state into many small-amplitude coherent states and trying to eliminate one of the phases first, and then use the optimal method to distinguish the remaining two possibilities.
So, first make $M$ copies (with $M\rightarrow \infty$)
of the state $|\alpha e^{i\phi}/\sqrt{M}\rangle$.
Then keep feeding copies in the three phase elimination setups until one of them gives a positive result.  
Subsequently use the rest of the light to elinate the last phase. The probability of success is
\begin{eqnarray}
P_{BS}^{(3)}&=&\lim_{M\rightarrow\infty} \sum_{k=0}^{M/3} 2e^{-6k|\alpha|^2/M}\left(1-e^{-3|\alpha|^2/M}\right)\nonumber\\&&\times
\left(1-e^{-3|\alpha|^2(1-3k/M)/2}\right)\nonumber\\
&=&1+3e^{-2|\alpha|^2}-4e^{-3|\alpha|^2/2}.
\end{eqnarray}
The first line gives the probability to produce a photodetection event after the $k$th try in one of the two photodetectors that possibly could click, the second line gives the probability of USD between the two remaining phases $\pi/2$ apart, with the light remaining at that point, with intensity $|\alpha|^2(1-3k/M)$.
The function $P_{BS}^{(3)}$ is plotted in Fig.~\ref{3} along with $P_D^{(3)}$ as a function of photon number $|\alpha|^2$.
For small amplitudes $\alpha$ the minimum coefficient determining the optimum USD probability according to (\ref{min}) is always $|c_{N-1}|^2$.
For $N=3$ we get $P_{D}^{(3)}\approx 3|\alpha|^4/2$, and comparing this to $P_{BS}^{(3)}\approx 3|\alpha|^4/2$ shows our beamsplitter scheme is in fact optimal for small photon numbers, but the simple scheme only reaches $\tilde{P}_{BS}^{(3)}\approx |\alpha|^4$.

\subsection{$N=4$}\label{N=4}
The case of $N=4$ symmetric coherent states is particularly relevant for QKD: As mentioned in the Introduction the 4 signal states in standard implementations of the BB84 protocol \cite{bb84} using phase encoding are symmetric coherent states. Because of its relevance we describe the measurement in more detail.

\begin{figure}\leavevmode
\epsfxsize=8cm \epsfbox{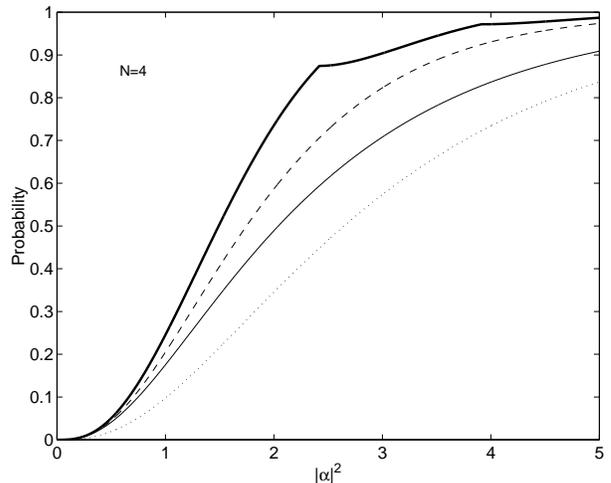} \caption{USD for $N=4$ symmetric coherent states. The thick solid curve gives the optimum success probability $P_D^{(4)}$, 
the dotted curve corresponds to a simple beamsplitter measurement $\tilde{P}_{BS}^{(4)}$, and the dashed curve describes a more complicated version of the beamsplitter setup $P_{BS}^{(4)}$.
The thin solid curve gives $P_{{\rm pol}}^{(4)}$, corresponding to the related USD measurement on polarization (see text).}
\label{4} \end{figure}

But first, we consider the simple scheme for the case $N=4$ consisting of splitting the coherent state into 4 equal parts $|\alpha e^{i\phi}\rangle/2$ and feeding these 4 states into 4 phase elimination setups. The probability to detect photons in 3 of those measurements is
\begin{eqnarray}
\tilde{P}_{BS}^{(4)}&=&
\left(1-e^{-|\alpha|^2/2}\right)^2\left(1-e^{-|\alpha|^2}\right),
\end{eqnarray}
which is plotted in Fig.~\ref{4}.

The full measurement procedure that can be used in BB84 and that generalizes the near-optimum measurement from the preceding subsection, consists of 4 steps:
\begin{enumerate}
\item Split the original state into $M$
copies of the state $|\alpha e^{i\phi}/\sqrt{M}\rangle$. The optimum procedure requires taking the limit $M\rightarrow \infty$ but any large value of $M$ will be sufficient in practice.\\
\item Choose randomly one of two ``bases'' (as in Section \ref{N=2}), either $\phi=0,\pi$ or $\phi=\pi/2,3\pi/2$, and perform phase measurements in the chosen basis on the subsequent copies until a photon is detected.
The probability to detect a photon in exactly $k+1$ out of a possible $M$ tries is
\begin{equation}
P_1^{(k)}=e^{-2k\alpha^2/M}(1-e^{-2\alpha^2/M}).
\end{equation}
The total probability to succeed in finding a photon is
\begin{equation}
P_1=1-e^{-2\alpha^2}.
\end{equation}
This probability is larger than the probability for a photon to be found in the original signal state, simply because the measurement procedure itself doubles the number of photons on average. If one insists on using classical intuition, however, it is surprising that in half of the cases the receiver manages to measure the phase setting used by the sender no photons would have been found in the signal state.
\item If a photon was detected in one basis in the previous step, then perform phase measurements in the other basis on the remaining copies until a photon is detected. 
If in the previous step exactly $k+1$ copies were used, the probability to detect a photon after exactly $m+1$ out of the remaining $M-k-1$ tries is 
\begin{equation}
P_2^{(m)}=e^{-2m\alpha^2/M}(1-e^{-2\alpha^2/M}).
\end{equation}
The total probability to detect at least 2 photons in total (one in this step, one in the previous) is
\begin{eqnarray}
P_2&=&\sum_{k=0}^{M-1} e^{-2k\alpha^2/M}(1-e^{-2\alpha^2/M})\nonumber\\
&&\times (1-e^{-2(M-k-1)\alpha^2/M})\nonumber\\
&=&1-e^{-2\alpha^2}-Me^{-2\alpha^2}\big(e^{2\alpha^2/M}-1\big).
\end{eqnarray}
In the limit of $M\rightarrow\infty$ this reduces to
\begin{equation}
P_2\rightarrow 1-e^{-2\alpha^2}-2\alpha^2e^{-2\alpha^2}.
\end{equation}
For small $\alpha$, $P_2\approx 2\alpha^4$, which is larger by a factor of 4  than the probability to find at least 2 photons in the original state for small $\alpha$.
With 50\% probability the {\em classical} bit values corresponding to the two outcomes obtained in this and the previous step will coincide. In that case, Bob knows which {\em classical} bit Alice sent without having to exchange basis information (in this sense this is like the B92 protocol \cite{B92}, and as such probably secure). If the two classical bit values obtained by Bob are different, then only Alice will have to reveal what basis she used for Bob to know what state Alice had sent. Of course if in such a case the eavesdropper detected two photons as well, this would not be secure. However, the fact that Bob detected 2 photons does not imply that Eve detected two photons. 
\item If a photon was detected in the previous step, take the remainder of the light and feed equal amounts in two phase elimination setups corresponding to the two remaining phases. The probability to detect another photon, and thus to unambiguously determine the quantum state, is
\begin{equation}
P_3=1-e^{-(M-k-m)\alpha^2/M}.
\end{equation}
The total USD probability is then
\begin{eqnarray}
P_{BS}^{(4)}&=&\sum_{k=0}^{M-1} e^{-2k\alpha^2/M}\big( 1-e^{-2\alpha^2/M} \big)
\nonumber\\ &&\times \sum_{m=0}^{M-k-2} e^{-2m\alpha^2/M}\big( 1-e^{-2\alpha^2/M} 
\big)\nonumber\\ &&\times  \big(1-e^{-|\alpha|^2(1-(k+1)/M-(m+1)/M)}\big),
\end{eqnarray}
which in the limit $M\rightarrow\infty$ simplifies to
\begin{equation}
P_{BS}^{(4)}\rightarrow 1+3e^{-2|\alpha|^2}+2|\alpha|^2e^{-2|\alpha|^2}-4e^{-|\alpha|^2}.
\end{equation}
This function is plotted in Fig.~\ref{4} together with the optimum $P_D^{(4)}$ as a function of photon number $|\alpha|^2$.
For small $|\alpha|^2$, the more complicated scheme is optimal, 
$P_{BS}^{(4)}\approx 2|\alpha|^6 /3$, but 
the simple scheme only reaches $\tilde{P}_{BS}^{(4)}\approx |\alpha|^6/4$.
\end{enumerate}
To conclude this subsection, we consider the case
where polarization is the degree of freedom used to encode information.
As mentioned before, the quantum states of the signals are mixed states. More precisely, the 4 signal states can be represented by 
density matrices of the form
\begin{eqnarray}
\rho_k=\int \frac{{\rm d}\phi}{2\pi} |\alpha/\sqrt{2} e^{i\phi}\rangle\langle \alpha/\sqrt{2} e^{i\phi}|\nonumber\\
\otimes
|\alpha/\sqrt{2} e^{i(\phi+\phi_k)}\rangle\langle \alpha/\sqrt{2} e^{i(\phi+\phi_k)}|,
\end{eqnarray}
with $\phi_k=2\pi k/4$ for $k=0\ldots 3$ and where the two modes correspond to two orthogonal polarizations.
One could say the phase of the signal states is defined relative to a reference pulse of amplitude $\alpha/\sqrt{2}$ with the effective signal states having the same amplitude.

The measurement procedure for phase encoding considered above requires for the first two photodetection events only reference pulses of the same amplitude as the signal. This part of the procedure thus carries directly over to the polarization case. It is only for the detection of the third photon, the part completing the USD measurement, that one needs to know the absolute phase of the signal states. This part, therefore, cannot be used for the polarization case. Instead, 
Ref.~\cite{norbert} gives an ingenious argument to derive the optimum USD measurement for this case:
produce a state $|\alpha\rangle$ of a mode with polarization $\vec{\epsilon}$ orthogonal to that of the unknown state (call it $\vec{\epsilon'}$). The combined state can be written as a coherent state with amplitude $\sqrt{2}\alpha$ with an unknown polarization
$(\vec{\epsilon}+e^{i\phi}\vec{\epsilon'})/\sqrt{2}$. Then perform a non-demolition measurement of the photon number in that state and then perform the optimum USD measurement of polarization
given the number of photons found. Such a measurement only exists when one found 3 or more photons and the resulting probability of successful discrimination between the 4 phases is
\begin{equation}
P_{{\rm pol}}^{(4)}=1-e^{-2|\alpha|^2}\big( \sqrt{2}\sinh (\sqrt{2}|\alpha|^2) +2\cosh (\sqrt{2}|\alpha|^2) -1 \big),
\end{equation}
which is plotted in Fig.~\ref{4}.
Since this measurement would require a quantum-non-demolition measurement of photon number and an unspecified optimum USD measurement of polarization on Fock states it may not be
implemented using just linear optics and photodetectors.
\subsection{$N>4$}\label{Ngg4}
It is easy to generalize the simple schemes of Subsections \ref{N=3} and \ref{N=4} to $N>4$.
Make $N$ copies of the state $|\alpha e^{i\phi}/\sqrt{N}\rangle$ and test each of them for one
of the phases $\phi_k=2\pi k/N$. The probability to detect at least one photon in $N-1$ of those
measurements is
\begin{equation}
\tilde{P}_{BS}^{(N)}=\prod_{k=1}^{N-1} \big( 1-e^{-|\alpha|^2/N |e^{2\pi ik/N}-1|^2} \big)
\end{equation}
For small $|\alpha|^2$ this expression reduces to \cite{note2}
\begin{eqnarray}
\tilde{P}_{BS}^{(N)}&\approx& 
\frac{|\alpha|^{2(N-1)}}{N^{N-1}}\prod_{k=1}^{N-1} |e^{2\pi ik/N}-1|^2=
\frac{|\alpha|^{2(N-1)}}{N^{N-3}}.
\end{eqnarray}
The optimal USD probability scales as
\begin{equation}\label{scale}
P_{D}^{(N)}\approx \frac{N|\alpha|^{2(N-1)}}{(N-1)!}.
\end{equation}
This shows that the simple scheme is far from optimal for large $N$: For small amplitudes $\alpha$   the ratio of the two probabilities scales as $\tilde{P}_{BS}^{(N)}/P_{D}^{(N)}\propto e^{-N}$ for large $N$.

One straighforward generalization of the more complicated beamsplitter schemes of the previous subsections consists of first making $M$ copies of $|\alpha e^{i\phi}/\sqrt{M}\rangle$, then use $N$ phase elimination setups until one of those setups detects a photon after which the remaining $N-1$ phases are tested, etc. 
Define $\theta_m$ to be the phase eliminated at the $m$th step. With the further definitions
\begin{eqnarray}
A_m&\equiv& |e^{i\theta_m}-1|^2\nonumber\\
A_{\geq m}&\equiv&\sum_{n=m}^{N-1}A_n\nonumber\\
s_m&=&\frac{M-\sum_{j=1}^{m-1}(N-j+1)k_j}{N-m+1},
\end{eqnarray} 
the probability to succeed in USD is then
\begin{eqnarray}
P_{BS}^{(N)}=\lim_{M\rightarrow\infty}
\sum_{\{\theta_i\}=\pi\{\phi_i\}}
\sum_{k_1=0}^{s_1}\ldots \sum_{k_m=0}^{s_m}\ldots\sum_{k_{N-1}=0}^{s_{N-1}} 
\nonumber\\ e^{-k_m|\alpha|^2 A_{\geq m}/M}\big(1-e^{-A_m|\alpha|^2/M}\big)
\end{eqnarray}
where the first summation is over all the $(N-1)!$ different orders the phases $\phi_i, i=1,2\ldots (N-1)$ 
can be eliminated. This expression is not easy to evaluate analytically, but for small $|\alpha|$ we can expand in powers of $|\alpha|$ and evaluate the lowest-order nontrivial term. Starting from the last summation over $k_{N-1}$ and working backwards to the summation over $k_1$ we note that each summation gives rise to a factor
\begin{equation}
a_n=\left(\frac{n+2}{n+1}\right)^{n} \frac{1}{n} \frac{A_{N-n} |\alpha|^2}{M}
\end{equation}
at the $n$th step.  The final summation then yields another factor $s_1^{N-1}=(M/N)^{N-1}$ and not forgetting the factor $(N-1)!$ for the summation over all permutations of $N-1$ phases, finally leads to
\begin{equation}
P_{BS}^{(N)}\approx (N-1)! s_1^{N-1}\prod_{n=1}^{N-1}a_n=
\frac{N|\alpha|^{2(N-1)}}{(N-1)!}.
\end{equation}
Comparing this expression with (\ref{scale}) we see that this is in fact optimal for small $\alpha$ for any $N$.
\section{Conclusions}
We presented linear-optics implementations of unambiguous state discrimination measurements on $N$ symmetric coherent states, that is, coherent states with known amplitude $\alpha$ but an unknown phase chosen from one of $N$ equidistant values in the interval $[0,2\pi)$.
We discussed very simple but sub-optimal schemes for the cases of $N=2,3,4$ that can be very easily implemented, and we extended these to more complicated schemes requiring feedback that are optimal \cite{note5} for small amplitude coherent states, the relevant limit for optical implementations of Quantum Cryptography. 

Indeed, the set-up for $N=4$ can be fruitfully used by the receiving end in a BB84 \cite{bb84} Quantum Key Distribution protocol with faint laser pulses. When polarization encodes the information only part of the protocol can be used, when phase encdoing is used the full procedure applies.
If three photons in total are detected, the protocol is in fact a USD measurement, if only two photons are detected the receiver will know the bit sent by the sender in 50\% of the cases without having to exchange any basis information. If only a single photon is detected, the measurement is equivalent to the standard one required in a single-photon implementation of the BB84 protocol. 

\section*{Acknowledgements}It's a pleasure to thank Norbert L\"utkenhaus for many useful comments and discussions.


\begin{thebibliography}{99}
\bibitem{note2}I thank Eric Rains for the 2 milliseconds he needed to show that $\prod_{k=1}^{N-1} |e^{2\pi ik/N}-1|^2=N^2$ for integer $N$.
\bibitem{note5}Of course, the scheme is optimal only with perfect photodetectors (no dark counts, 100\% efficiency), perfect beamsplitters, perfect alignment, etc. etc. etc.
\end{thebibliography}
\end{document}